\documentclass{article}
\usepackage{spconf,amsmath,graphicx,hyperref}
\usepackage{siunitx}
\usepackage{booktabs,multirow}
\usepackage{color,soul}
\usepackage{tikz}
\usetikzlibrary{positioning,arrows.meta,calc}

\title{Word Timestamps and Speaker Attribution with a Non-Autoregressive LLM}
\name{Zvi Kons, Avihu Dekel, Hagai Aronowitz, Vishal Sunder, Ron
  Hoory
}
\address{IBM Research}

\begin{document}

\maketitle
\begin{abstract}
  Timestamps and speaker attribution are useful additions to speech
  recognition, creating a rich text transcript. This information can
  either be extracted during transcription or aligned to a given
  transcript. In this paper we present models that add timestamps and
  speaker information to a given transcript using a non-autoregressive
  LLM-based architecture. Compared to an
  autoregressive model built from similar components, the models are
  more accurate and annotate a
  given transcript one to two orders of magnitude faster.
  Compared to other models, our models achieve
  state-of-the-art timestamp accuracy and the best cpWER for speaker
  attribution.
\end{abstract}
\section{Introduction}
\label{sec:intro}

Granite-Speech is a family of automatic speech recognition (ASR)
models, of various sizes and features, released under a permissive
license. Among them is Granite-Speech-Plus
(GSP)\footnote{https://huggingface.co/ibm-granite/granite-speech-4.1-2b-plus},
an autoregressive (AR), LLM-based model for rich speech
transcription. Beyond plain ASR, it offers ASR with word-level
timestamps (TS) \cite{fan2026insyncadaptationspeechaware} and
speaker-attributed ASR (SAA) \cite{hagai2026saa}, in both cases adding
the extra information to the transcript as textual tags. Its downside
is that its AR nature, together with the additional tag tokens it must
generate, makes it slower than many other models.

Another relevant family member is
Granite-Speech-NAR\footnote{https://huggingface.co/ibm-granite/granite-speech-4.1-2b-nar}
\cite{dekel2026nle}. It uses components similar to GSP (audio encoder,
projector and LLM decoder), but the encoder output is first decoded by
a Connectionist Temporal Classification (CTC) head into a draft
transcript, and the LLM then acts as a non-autoregressive (NAR)
bidirectional editor that turns the draft into the final transcript in
a single pass, giving a very fast model without compromising accuracy.

In this paper we extend the NAR architecture to transcript annotation
with word-level timestamps and speaker attribution. The models take a
speech signal and its transcript as input, and annotate the transcript
with textual tags for either timestamps or speaker ID numbers. Using
the same training data as the AR models, they are more accurate and
annotate a given transcript far faster, and against other available
tools they achieve state-of-the-art results for TS and the best cpWER
for SAA.

\subsection{Related work}
\label{sec:related}

\noindent\textbf{Word level timestamps.}
Aligning a known transcript to audio is classically done by hybrid
forced aligners such as MFA \cite{mcauliffe2017montreal}, which rely
on language-specific lexicons and phoneme sets, or by aligners driven
by the frame-level posteriors of a CTC model
\cite{rastorgueva2023nfa}. Other systems derive timestamps from a
recognizer that was not trained to produce them, either by dynamic
time warping against a phoneme model \cite{bain2023whisperx} or by
selecting the attention heads that happen to encode alignment
\cite{yeh2025whisperaligner,zusag2024crisperwhisper}. A second approach
lets the recognizer emit timestamps itself as a by-product of
transcription: non-autoregressive CIF models predict token boundaries
during recognition \cite{shi2022achieving}, while AR speech-aware LLMs
emit time tags as ordinary text tokens
\cite{hu2025wordlevel,fan2026insyncadaptationspeechaware}, paying for
every tag with additional decoding steps and with no guarantee of
monotonicity. Closest to our work is LLM-ForcedAligner \cite{qwennar},
which also casts alignment as filling time slots in a given transcript;
we differ in predicting each timestamp as a small set of digit tokens
over a bidirectional encoding of the whole transcript, rather than as
an index into a single large time vocabulary.

\noindent\textbf{Speaker-attributed ASR.}
Serialized output training \cite{kanda2020serialized} established the
formulation we follow, in which speaker changes are written inline in
one token stream instead of being resolved by a separate diarization
stage; it remains the basis of recent end-to-end SAA systems
\cite{10389762}. The conventional alternative is a cascade, in which a
diarization pipeline \cite{pyannoteBredin23,pyannotePlaquet23} runs
alongside ASR and the two are merged. LLMs have also been used to
repair such merges after the fact: DiarizationLM
\cite{wang2024diarizationlm} post-processes a diarized transcript with
an LLM and substantially reduces WDER.

\noindent\textbf{Non-autoregressive decoding.}
NAR speech recognition trades the sequential context of AR decoding
for parallelism \cite{higuchi2021comparative}. We inherit this
property from the NAR editing architecture of \cite{dekel2026nle}, and
observe that annotation tasks suit it especially well: the tags are
conditionally independent given the transcript and the audio, so
little is lost by emitting them simultaneously.

\section{Model architecture}
\label{sec:arch}

%

\newcommand{\snowflake}[1][0.075]{%
  \begin{tikzpicture}[scale=#1,baseline=-0.55ex,line width=1.1pt,
                      draw=cyan!60!blue]
    \foreach \a in {0,60,120} {\draw (\a:1) -- (\a+180:1);}
    \foreach \a in {0,60,120,180,240,300} {
      \draw (\a:1) -- ++(\a+145:0.38);
      \draw (\a:1) -- ++(\a-145:0.38);
    }
  \end{tikzpicture}}
\newcommand{\flame}[1][0.075]{%
  \begin{tikzpicture}[scale=#1,baseline=-0.55ex]
    \fill[orange!88!red]
      (0,1.35) .. controls (0.40,0.72) and (0.98,0.35) .. (0.72,-0.30)
               .. controls (0.58,-0.72) and (-0.58,-0.72) .. (-0.72,-0.30)
               .. controls (-0.98,0.35) and (-0.40,0.72) .. (0,1.35) -- cycle;
    \fill[yellow!88!orange]
      (0.02,0.52) .. controls (0.30,0.10) and (0.40,-0.16) .. (0.28,-0.36)
               .. controls (0.16,-0.54) and (-0.24,-0.54) .. (-0.32,-0.32)
               .. controls (-0.42,-0.10) and (-0.20,0.18) .. (0.02,0.52) -- cycle;
  \end{tikzpicture}}
\newcommand{\strip}[1]{%
  \begin{tikzpicture}[baseline=-0.5ex,
      cell/.style={draw=black!65,line width=0.4pt,minimum width=1.4mm,
                   minimum height=2.9mm,inner sep=0pt},
      acell/.style={cell,fill=green!45!teal!45},
      wcell/.style={cell,fill=blue!55!black!70},
      hcell/.style={cell,fill=white},
      tcell/.style={cell,fill=orange!70}]
    \foreach \ty [count=\i from 0] in {#1} {\node[\ty] at (\i*1.7mm,0) {};}
  \end{tikzpicture}}
\newcommand{\vstrip}[1]{%
  \begin{tikzpicture}[baseline=-0.5ex,
      cell/.style={draw=black!65,line width=0.4pt,minimum height=1.4mm,
                   minimum width=2.9mm,inner sep=0pt},
      acell/.style={cell,fill=green!45!teal!45},
      wcell/.style={cell,fill=blue!55!black!70},
      hcell/.style={cell,fill=white},
      tcell/.style={cell,fill=orange!70}]
    \foreach \ty [count=\i from 0] in {#1} {\node[\ty] at (0,-\i*1.7mm) {};}
  \end{tikzpicture}}

\newcommand{\lgnd}[1]{%
  \raisebox{\dimexpr0.62pt+0.18ex\relax}{%
    \tikz{\node[draw=black!65,line width=0.4pt,
      minimum height=1.9mm,minimum width=2.9mm,inner sep=0pt,fill=#1] {};}}}
\begin{figure}[t]
  \centering
  \resizebox{\columnwidth}{!}{%
  \begin{tikzpicture}[
      font=\small,
      >={Stealth[length=4pt,width=3.2pt]},
      node distance=3.2mm and 4.2mm,
      blk/.style={rounded corners=2pt,draw=black!70,line width=0.5pt,
                  align=center,inner sep=2.4pt,minimum height=7.6mm},
      prj/.style={blk,fill=blue!20},
      enc/.style={blk,fill=green!30!teal!26},
      tok/.style={blk,fill=orange!28},
      llm/.style={blk,fill=cyan!20},
      pp/.style={blk,fill=black!7},
      lbl/.style={font=\small,align=center,inner sep=1.2pt},
      txt/.style={font=\small\itshape,text=violet!72!black,align=center},
      ar/.style={->,draw=black!75,line width=0.5pt},
    ]

    \node[lbl,align=left,anchor=west] (wave) at (0,0) {audio};

    \node[enc,right=of wave,text width=15mm] (enc)
      {Conformer\\encoder\,\snowflake};

    \node[prj,right=of enc,text width=14mm] (prj) {Q-Former\\projector\,\flame};

    \coordinate (intxtx) at ($(wave.west)-(0,13mm)$);
    \node[lbl,align=left,anchor=west] (intxt) at (intxtx)
      {input\\transcript};

    \node[tok,right=of intxt,text width=17mm] (tok) {tokenize $+$\\placeholders};

    \coordinate (mid) at ($(prj.south)!0.5!(tok.north)$);

    \coordinate (seqx) at ($(prj.east)+(7mm,0)$);
    \node[anchor=south,inner sep=0pt] (aseq) at (seqx |- mid)
      {\vstrip{acell,acell,acell,acell,acell,acell}};
    \node[anchor=north,inner sep=0pt,yshift=0.3mm] (tseq) at (aseq.south)
      {\vstrip{wcell,hcell,wcell,hcell,wcell,hcell}};

    \coordinate (llmx) at ($(aseq.east)+(6mm,0)$);
    \node[llm,text width=14mm,minimum height=17mm,anchor=west] (llm)
      at (llmx |- mid)
      {Bidi\\LLM\\[1.5pt]LoRA\,\flame};

    \coordinate (outx) at ($(llm.east)+(6mm,0)$);
    \node[anchor=south,inner sep=0pt] (aout)
      at (outx |- mid)
      {\vstrip{acell,acell,acell,acell,acell,acell}};
    \node[anchor=north,inner sep=0pt,yshift=0.3mm] (tout) at (aout.south)
      {\vstrip{wcell,tcell,wcell,tcell,wcell,tcell}};
    \coordinate (ppx) at ($(aout.east)+(6mm,0)$);
    \node[pp,text width=17mm,anchor=west] (pp)
      at (ppx |- mid) {post-\\processing};
    \coordinate (outtxtx) at ($(pp.east)+(5mm,0)$);
    \node[txt,anchor=west] (out) at (outtxtx |- mid) {rich\\transcript};

    \coordinate (lgx) at ($(intxt.west)!0.5!(pp.east)$);
    \coordinate (lgy) at ($(tseq.south)-(0,4.5mm)$);
    \node[lbl,anchor=north] (legend) at (lgx |- lgy)
      {\lgnd{green!45!teal!45}\,acoustic\quad\lgnd{blue!55!black!70}\,transcript\quad\lgnd{white}\,placeholder\quad\lgnd{orange!70}\,SAA\,/\,TS tag};

    \draw[ar] (wave.east) -- (enc.west);
    \draw[ar] (enc.east) -- (prj.west);
    \draw[ar] (intxt.east) -- (tok.west);

    \draw[ar,rounded corners=1.6pt] (prj.east)
      -- ($(prj.east)!0.55!(prj.east -| aseq.west)$)
      |- (aseq.west);
    \draw[ar,rounded corners=1.6pt] (tok.east)
      -- ($(tok.east)!0.55!(tok.east -| tseq.west)$)
      |- (tseq.west);

    \coordinate (seqout) at ($(aseq.east |- mid)+(0.6mm,0)$);
    \coordinate (outin)  at ($(aout.west |- mid)-(0.6mm,0)$);
    \coordinate (outout) at ($(aout.east |- mid)+(0.6mm,0)$);
    \draw[ar] (seqout) -- (llm.west);
    \draw[ar] (llm.east) -- (outin);
    \draw[ar] (outout) -- (pp.west);
    \draw[ar] (pp.east) -- (out.west);

  \end{tikzpicture}}
  \caption{Overview of the NAR annotation model. The frozen conformer
    encoder produces acoustic embeddings and a trained Q-Former projector
    maps them into the LLM space; the given transcript is tokenized with
    placeholder slots inserted after each word. The two are concatenated
    into the single sequence that the bidirectional, LoRA-adapted LLM
    consumes, returning it with every placeholder replaced by its SAA or
    TS tag in a single pass. \snowflake~frozen, \flame~trained.}
  \label{fig:arch}
\end{figure}
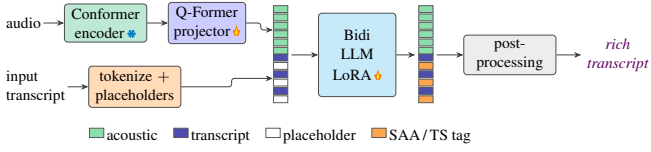

Our model follows a common ASR architecture: a pre-trained audio
encoder connected by a projector to an LLM module
\cite{ma2026slam,saon2025granite} (Figure~\ref{fig:arch}). The encoder
is a multi-layer conformer pre-trained on ASR-only tasks with CTC
loss. Since we found that its output contains very little speaker
information, we concatenate the output of an intermediate layer to
that of the final one \cite{hagai2026saa}.

The input transcript is tokenized and placeholder tokens are inserted
after each word (see below), and the LLM must replace them with the SAA
or TS tags in a single NAR pass, leaving all other tokens unchanged. It
is based on
Granite-4.0-1B-base\footnote{https://huggingface.co/ibm-granite/granite-4.0-1b-base}
and modified to use bidirectional attention, so that every position can
attend to all the audio, text and tag tokens. In both tasks the tags
are selected by $\operatorname{argmax}$ over the logits of the tokens
for the numbers 0--999, and the output is post-processed into a rich
text transcript with tags in a format similar to GSP's.

\subsection{SAA tags}
The speaker tag is a speaker ID number, with speakers numbered by
their order of appearance in the audio sample. Our tokenizer encodes
every number from 0 to 999 as a single token, so a single-token
placeholder is inserted after every word and the model is trained to
replace it with that word's speaker ID. The output is post-processed
into a transcript with turn-level speaker tags of the form
\texttt{"[Speaker 1]: speaker 1 turn \ldots [Speaker 2]: speaker 2 turn
\ldots "}. We assume each word has a single speaker; when speakers
overlap, each speaker's words appear separately in the transcript.

\subsection{Timestamp tags}
In the timestamps task we find the start and end time of each
word. Each timestamp $t$ is encoded by two integers in the range
0--999: $f \bmod 1000$ and $\lfloor f/1000 \rfloor$, where
$f=\lfloor t/\qty{10}{\milli\second} \rfloor$ is the frame number.
Each integer takes one token, for a total of four tokens per word.
A single token could have covered the whole timestamp range (as in
\cite{qwennar}), but that would require either training a new output
head or having the LLM assign the token a meaning very different from
its original one.

Accordingly, four placeholder tokens are inserted after each word and
the model is trained to replace them with the start and end
timestamps. Post-processing first restores monotonicity, moving any
timestamp smaller than the previous or larger than the following one to
lie between its neighbors, and then converts the output to textual tags
for each word's end time (e.g.\ \texttt{"hello [T:25] world [T:42]"}),
while the start times are used to identify silences between words.

\section{Experiments}
\subsection{Training datasets}
\label{sec:datasets}

Word-level timestamping capabilities are achieved using a combination
of publicly available speech corpora: LibriSpeech, MLS (en, fr, de,
pt, es), CommonVoice (en, fr, de, pt, es), VoxPopuli (en, fr, de, es),
AMI-IHM, Switchboard, TIMIT and YODAS
\cite{Librispeech,pratap2020mls,ardila2020common,wang2021voxpopuli,kraaij2005ami,godfrey1992switchboard,lyons1993darpa,li2023yodas}.
For AMI-IHM, Switchboard and TIMIT we use the available timestamp
annotations; for all other datasets we obtain word-level alignments
with the Montreal Forced Aligner (MFA) \cite{mcauliffe2017montreal},
by the same process used for GSP training. For SAA we used Fisher,
CallHome, AMI-SDM and NaturalVoices
\cite{fisher2004,callhome,kraaij2005ami,Salman_2024} as in
\cite{hagai2026saa}.

We also generated longer audio samples by concatenating random samples
from the above datasets on the fly during training, so we can insert
random silences or create small overlaps between segments. For TS we
concatenated already-aligned samples and simply shifted their
timestamps; for SAA we randomly interleaved 2 to 6 speakers from
datasets with speaker labels (e.g.\ MLS). Since at inference the input
transcript is likely to come from an ASR system and contain recognition
errors, we also apply random word deletions to the input transcript
during training. In total we used about 81K hours of TS and 51K hours
of SAA training data.

\subsection{Training}

We train two models, one for TS and one for SAA, each only on the
corresponding datasets. Both share exactly the same architecture,
differing only in the transcript pre-processing (placeholder insertion)
and the post-processing of the output tokens.

The encoder is a stack of 16 conformer blocks initialized from a
CTC-ASR pre-trained model \cite{saon2025granite}, with the output of
the 3rd block stacked with that of the last one, so each frame is a
2048-dimensional vector. The projector is a single-layer window query
transformer (Q-Former) operating on blocks of 15 acoustic embeddings,
downsampled by a factor of 5 using 3 trainable queries per block,
yielding a \qty{10}{\hertz} frame rate. The LLM uses the
Granite-4.0-1B-base dense architecture and was initialized from an
intermediate checkpoint of that model.

During training we adapted only the projector weights and used LoRA
with a rank of 128 to adjust the LLM, so of the model's \num{2.11}\,B
parameters only \num{205}\,M are trained. We used the AdamW optimizer
with a cross-entropy loss computed only over the generated TS or SAA
tokens, on a cluster of 32 H100 GPUs for about 4 days.

\subsection{Scoring}
\label{sec:scoring}

We measure SAA accuracy with both the word diarization error rate
(WDER) and the concatenated minimum-permutation word error rate (cpWER)
\cite{watanabe2020chime6}. WDER is the rate of words attributed to the
wrong speaker; comparing the transcript to the ground truth (GT), we
count only words that can be matched (i.e.\ no insertions or
deletions). For cpWER the words of each speaker are concatenated and
the word errors are summed over speakers. In both cases we search for
the best speaker permutation, as the numbering might change. Because
WDER ignores insertions and deletions, it does not penalize a system
for words it failed to transcribe at all, which favours systems that
transcribe less; cpWER accounts for them, and for a system given a
transcript, such as ours, it is almost entirely attribution error. For
the timestamps task we report the accumulated averaging shift (AAS)
metric \cite{shi2022achieving}: the mean absolute error in milliseconds
between the predicted and the reference word end times, again computed
only over matched words.

Both tasks require an input transcript. Training uses the GT text, but a
real-world system must also cope with recognition errors, so we test
the model with two ASR transcripts as well: the GSP transcript (after
removing all tags), which allows a direct comparison with the AR model,
and that of Whisper-large-v3-turbo \cite{radford2022whisper}, whose
WER is generally higher than GSP's.

\subsection{Timestamps results}
\label{sec:ts_results}

Table~\ref{tab:ts} reports the AAS on eight English and eight
multilingual test sets, together with four external systems: the NAR
forced aligner Qwen3-FA \cite{qwennar}, which like our model aligns a
given transcript, and the three ASR systems CrisperWhisper
\cite{zusag2024crisperwhisper}, Canary-v2 \cite{sekoyan2025canary} and
WhisperX \cite{bain2023whisperx}, which emit timestamps together with
the transcript. Our model has the lowest AAS of all systems on every
one of the sixteen test sets, and stays ahead of all the external
systems even when the input transcript is not the GT. Against GSP it
reduces the average AAS by 32\% for English and 55\% for the other
languages; given the same GSP input transcript the reductions are 34\%
and 56\%. It is also insensitive to transcript quality: its results
with the GSP and Whisper text are very close to those with the GT.

We also evaluated a recent training-free aligner that reads word
boundaries out of the cross-attention maps of a frozen Whisper model
\cite{yeh2025whisperaligner}. Its average AAS is
\qty{77.9}{\milli\second} on English and \qty{72.4}{\milli\second} on
the multilingual sets --- better than WhisperX and Canary-v2 on both,
but worse than every other system in Table~\ref{tab:ts}.

\begin{table}[t]
  \centering
  \caption{Word-level timestamps, AAS ($\downarrow$\si{\milli\second}).}
  \label{tab:ts}
  \setlength{\tabcolsep}{3.2pt}
  \sisetup{table-format=3.1}
  \resizebox{\columnwidth}{!}{%
  \begin{tabular}{l S S S S S S S S}
    \toprule
     & {Qwen3} & {Crisper} & {Canary} & {Whisper} & {GSP} & \multicolumn{3}{c}{Our-NAR} \\
    \cmidrule(l){7-9}
    Test set & {-FA} & {Whisper} & {-v2} & {X} & & {GT} & {GSP} & {Whis.} \\
    \midrule
    AMI-IHM     &  48.1 &  55.7 & 114.5 & 107.1 & 43.6 & \bfseries 29.2 & 30.9 & 46.4 \\
    AMI-SDM     &  82.5 &  64.3 & 116.5 & 150.2 & 68.1 & 61.2 & \bfseries 50.5 & 66.9 \\
    LS clean    &  27.8 &  35.9 &  78.6 &  71.7 & 11.5 & \bfseries  9.5 & 11.6 & 11.5 \\
    LS other    &  29.3 &  40.1 &  75.7 &  72.0 & 14.5 & \bfseries 11.7 & 14.3 & 14.2 \\
    VoxPopuli   &  41.0 &  47.2 &  96.4 &  78.8 & 75.6 & 25.9 & \bfseries 21.6 & 22.4 \\
    CommonVoice &  48.4 &  97.4 & 102.8 &  91.2 & 42.6 & 31.4 & \bfseries 30.4 & 31.7 \\
    MLS         &  34.3 &  46.4 &  82.6 &  79.2 & 24.0 & 17.7 & \bfseries 17.3 & 20.2 \\
    TIMIT       &  29.9 &  42.7 &  77.7 &  63.6 & 21.8 & \bfseries 20.1 & 22.2 & 22.2 \\
    \midrule
    En avg      &  42.7 &  53.7 &  93.1 &  89.2 & 37.7 & 25.8 & \bfseries 24.8 & 29.4 \\
    \midrule
    MLS-fr      &  38.1 &  35.6 &  77.6 & 117.3 & 44.5 & \bfseries  8.7 &  8.9 &  9.5 \\
    MLS-es      &  27.0 &  28.0 &  75.2 &  84.7 & 21.1 & \bfseries  9.3 & \bfseries  9.3 &  9.7 \\
    MLS-de      &  31.2 &  31.2 &  70.9 & 132.2 & 42.1 & \bfseries  8.4 &  9.2 & 10.5 \\
    MLS-pt      &  26.3 &  36.8 &  79.8 &  75.0 & 45.7 & \bfseries 15.1 & 15.7 & 15.4 \\
    CV-fr       &  30.3 &  62.9 &  84.6 & 104.2 & 18.5 & 14.7 & \bfseries 14.0 & 15.1 \\
    CV-es       &  40.0 &  58.9 &  84.4 &  88.1 & 19.2 & 14.9 & \bfseries 14.3 & 14.8 \\
    CV-de       &  29.4 &  60.9 &  86.6 & 126.8 & 19.4 & 13.5 & \bfseries 13.2 & 14.2 \\
    CV-pt       &  34.2 &  83.8 &  93.5 &  79.5 & 23.8 & 19.9 & \bfseries 19.4 & 19.6 \\
    \midrule
    ML avg      &  32.1 &  49.8 &  81.6 & 101.0 & 29.3 & 13.1 & \bfseries 13.0 & 13.6 \\
    \bottomrule
  \end{tabular}}
\end{table}

\subsection{SAA results}
\label{sec:saa_results}

Table~\ref{tab:saa} reports results on three conversational test sets:
Fisher, CallHome English and AMI-SDM. Following \cite{hagai2026saa},
each score is the average over the \qty{2}{\minute} and
\qty{5}{\minute} segmentations of the test set.

VibeVoice-ASR\footnote{https://huggingface.co/microsoft/VibeVoice-ASR-HF}
is an end-to-end speaker-attributed ASR model that predicts speaker
labels jointly with the transcript (similar to GSP). It has the lowest
WDER on average, and the lowest of any system on AMI-SDM and CallHome,
but the NAR model is better on Fisher. However, WDER is not comparable
between systems with different deletion rates. We therefore also look at
cpWER, and there the ordering reverses: the NAR model is best on every
test set, because the words VibeVoice does not transcribe are now
penalized rather than excluded, and with the GSP transcript it is
comparable to VibeVoice. On top of that we measured the VibeVoice RTFx
at 19, about 2.5 times slower even than GSP (Sec.~\ref{sec:speed}), so
combining the NAR tagger with a fast ASR gives a pipeline as accurate as
VibeVoice but much faster.

We also compare against a cascaded baseline representing the
conventional approach: pyannote~3.1
\cite{pyannoteBredin23,pyannotePlaquet23} diarization followed by GSP
ASR on each resulting segment independently. The NAR model beats the
cascade on WDER on every test set, and on cpWER the cascade is the worst
system on average, probably because of the deletions caused by missed
segments. Overall the NAR model is robust to the different input
transcripts, and with the GSP transcript it is in most cases better than
GSP itself.

\begin{table}[t]
  \centering
  \caption{Speaker-attributed ASR.}
  \label{tab:saa}
  \sisetup{table-format=2.2}
  \resizebox{\columnwidth}{!}{%
  \begin{tabular}{l S S S S S S}
    \toprule
     & {VibeVoice} & {Cascade} & {GSP} & \multicolumn{3}{c}{Our-NAR} \\
    \cmidrule(l){5-7}
    Test set & {ASR} & {pyan.+GSP} & & {GT} & {GSP} & {Whisper} \\
    \midrule
    \multicolumn{7}{l}{\itshape WDER$\downarrow$ (\%)} \\
    Fisher            &  1.01 &  3.80 &            0.87 &  \bfseries 0.49 &  0.99 &  2.66 \\
    CallHome          & \bfseries  1.30 &  8.07 &     2.43 &           1.58 &  1.75 &  5.59 \\
    AMI-SDM           & \bfseries  5.93 & 13.76 &   15.46 &          10.10 & 10.53 & 11.44 \\
    Average           & \bfseries 2.75 &  8.54 &       6.25 &  4.05 &  4.43 &  6.56 \\
    \midrule
    \multicolumn{7}{l}{\itshape cpWER$\downarrow$ (\%)} \\
    Fisher            & 21.71 & 27.59 & 18.36 &  \bfseries 0.92 & 18.55 & 23.91 \\
    CallHome          & 19.70 & 35.11 & 20.65 &  \bfseries 2.90 & 19.69 & 26.68 \\
    AMI-SDM           & 28.42 & 46.67 & 41.70 & \bfseries 17.85 & 35.44 & 37.83 \\
    Average           & 23.28 & 36.46 & 26.91 & \bfseries  7.22 & 24.56 & 29.47 \\
    \midrule
    Average WER$\downarrow$ (\%)   & 21.84 & 29.93 & 21.26 & {--} & 21.28 & 22.24 \\
    \bottomrule
  \end{tabular}}
\end{table}

\subsection{Decoding speed}
\label{sec:speed}

With the same total number of parameters, the NAR model is also much
faster than the AR GSP model, since it requires only one pass of the
LLM. Table~\ref{tab:rtfx} compares the decoding speed of the models as
the inverse real-time factor (RTFx), the ratio of decoded audio duration
to wall-clock decoding time, for GSP in SAA and TS modes, for the NAR
model tagging a given transcript, and for a pipeline of GSP in ASR mode
followed by NAR tagging. All models ran on a single GPU in exclusive
mode, timing the decoding loop only, and we pool over the test sets on
which the models decoded exactly the same samples.
On the tagging pass itself the NAR model is \num{30} times faster than
GSP on SAA and \num{197} times faster on timestamps; the two tasks
differ because timestamps use four tag tokens per word against one for
SAA. NAR speed is also far less sensitive to input length: on the
Fisher segmentations its RTFx stays between \num{1374} and \num{1816}
from \qty{10}{\second} to \qty{300}{\second} clips, whereas GSP peaks
at \qty{30}{\second} and falls to \num{19} on the longest ones.

The NAR pipeline is dominated by the ASR speed. With a faster ASR such
as Granite-Speech-NAR, whose reported RTFx is about \num{1800}
\cite{dekel2026nle}, we would get about \num{800} and \num{600} for SAA
and TS respectively.

\begin{table}[t]
  \centering
  \caption{Tests sets size and decoding speed as RTFx$\uparrow$. GSP, NAR tagging and ASR+NAR.}
  \label{tab:rtfx}
  \begin{tabular}{l S[table-format=2.0] S[table-format=4.1] S[table-format=2.1] S[table-format=4.0] S[table-format=2.1]}
    \toprule
    Task & {Sets} & {Audio} & {GSP} & {NAR} & {ASR+NAR} \\
    \midrule
    SAA        & 43 &  \qty{942.7}{\hour} & 48.3 & 1454 & 53.3 \\
    Timestamps & 15 &  \qty{158.0}{\hour} &  4.6 &  902 & 23.9 \\
    \bottomrule
  \end{tabular}
\end{table}

\subsection{Causal vs.\ bidirectional attention}
\label{sec:attention}

To verify our choice of bidirectional attention for the LLM, we trained
the same models with causal (unidirectional) attention, where each tag
sees the audio tokens and those of the preceding words but not of the
following ones --- somewhat like the AR model, except that the AR model
sees committed discrete tokens whereas the NAR model sees only
representations of prior words. Table~\ref{tab:atten_short} shows that
the impact on SAA is not significant, while for timestamp tags causal
attention causes a large degradation.

\begin{table}[t]
  \centering
  \caption{Causal vs.\ bidirectional LLM attention, averaged over the
    test sets of Tables~\ref{tab:saa} and~\ref{tab:ts} with the GT
    input transcript.}
  \label{tab:atten_short}
\begin{tabular}{l l S[table-format=2.2] S[table-format=2.2] S[table-format=2.1,explicit-sign=+]}
  \toprule
  Task & Metric & {Bi-dir.} & {Uni-dir.} & {Rel.\ [\%]} \\
  \midrule
  SAA    & WDER [$\downarrow$\%]                        & \bfseries 10.76 & 10.78 &  +0.2 \\
  TS, En & AAS [$\downarrow$\si{\milli\second}]         & \bfseries 25.8 & 33.0 & +27.9 \\
  TS, ML & AAS [$\downarrow$\si{\milli\second}]         & \bfseries 13.1 & 17.7 & +35.1 \\
  \bottomrule
\end{tabular}
\end{table}

\section{Conclusions}

We presented two new models for speaker attribution and word-level
timestamps, whose novelty is the use of an LLM to tag a given transcript
in a single non-autoregressive pass. Tested on a variety of test sets
against other available models, they outperform a similar AR model on
most tests even when tagging the same transcript; on TS our model is
better than all the others, and on SAA, when paired with an accurate
transcript, it achieves the best cpWER. Requiring only one LLM pass also
makes the models fast, so combining them with a fast ASR yields a very
fast rich-transcription pipeline, and because they work on a given
transcript, both SAA and TS tagging are easy to apply to the same audio.
Additional work is needed to merge both tasks into one model. We plan to
release the models under a permissive license in the near future as part
of the Granite-Speech family.


\section{Compliance with Ethical Standards}

No ethical approval was required since we used only existing
corpora, obtained and used under their respective licenses
and no new data was collected.

\let\oldthebibliography\thebibliography
\renewcommand{\thebibliography}[1]{%
  \oldthebibliography{#1}%
  \setlength{\itemsep}{0pt plus 0.3pt}%
  \setlength{\parsep}{0pt}%
  \setlength{\parskip}{0pt}%
  \setlength{\topsep}{0pt}%
}

\bibliographystyle{IEEEbib}
\bibliography{refs}

\end{document}